\documentclass[12pt]{iopart}
\usepackage[dvips]{graphicx}
\usepackage{epsf}
\usepackage{amsfonts}
\usepackage{amssymb}
\usepackage{latexsym}
 
\def\ltsima{$\; \buildrel < \over \sim \;$}
\def\simlt{\lower.5ex\hbox{\ltsima}}

\def\be{\begin{equation}}
\def\ee{\end{equation}}
\def\ba{\begin{eqnarray}}
\def\ea{\end{eqnarray}}
\def\ben{\begin{displaymath}}
\def\een{\end{displaymath}}

\def\p{\partial}

\begin{document}

\title[Race for the Kerr field]
{Race for the Kerr field}

\author{G. Dautcourt\footnote{
Max-Planck-Institut f\"ur Gravitationsphysik, Albert-Einstein-Institut \\
Am M\"uhlenberg 1, D-14476 Golm, Germany,~~  Email:~daut@aei.mpg.de}}

\begin{abstract}\baselineskip=12pt
Roy P. Kerr has discovered his celebrated metric 45 years ago,
yet the problem to find a generalization of the Schwarzschild metric
for a rotating mass was faced much earlier. 
Lense and Thirring, Bach, Andress, Akeley, Lewis, van Stockum and others 
have tried to solve it or to find an approximative solution at least. In 
particular Achilles Papapetrou, from 1952 to 1961 in Berlin,  was          
interested in an exact solution. He directed the author in the late autumn of 
1959 to work on the problem. Why did these pre-Kerr attempts fail? 
Comments based on personal reminiscences and old notes.
\end{abstract}

\baselineskip=15pt
\section{\label{sec:level1} Introduction}
The old Prussian Academy of Sciences in Berlin was certainly a good 
place after the Second World War to continue the research
of its most prominent former member, Albert Einstein.
Here Einstein had worked nineteen years and created his beautiful 
theory of gravitation. 
Here several attempts had  been undertaken to test the  
theory: Einstein's young coworker, the astronomer Erwin Finley Freundlich, 
had tried to see (with little success) gravitational redshift effects in 
astronomical objects like the Sun.
Freundlich was also active in several campaigns to observe Solar eclipses 
with the aim to verify the predicted displacement of stars seen near the Sun.  
And here Karl Schwarzschild, the director of the Astrophysical 
Observatory at Potsdam, a town at the outskirts of Berlin, had              
published in 1916 the first exact solution of Einstein's field 
equations, describing the exterior gravitational field of a
nonrotating spherical mass \cite{Schwarzschilda}. 
In view of the extreme nonlinearity of the equations, it appeared
almost as a miracle that exact solutions exist at all. 

The administrative buildings of the Academy 
(in the summer of 1946 reopened as "German Academy of Sciences") 
and of 
the Berlin university happened to be situated in the Eastern part of 
the divided city of Berlin, thus the Academy worked 
under the influence of the Soviet Military Administration and later the 
East German government. Other scientific 
institutions such as institutes of the former Kaiser-Wilhelm society
in Berlin-Dahlem and the 1948 founded Free University, also there, 
did belong to West Berlin, the sphere of influence of the Western Allies.
The borders between the two parts of Berlin were open until 1961, 
allowing at least
some personal contact between scientists of both sides. 
The director of the Academy since November 1946 was Josef Naas, 
a mathematician and member of the communist party, who was sent to a 
concentration camp in the time of the Nazi regime.
He and other officials of the Academy were interested in a 
continuation of research on Einstein's path at this traditional place.
In 1952 Achilles Papapetrou, a Greek scientist at the
Physics Department of the University of Manchester, was invited as Senior 
Researcher 
to the Academy's Research Institute for Mathematics, headed by Naas.
Papapetrou had started his academic career in solid state 
theory in the German town Stuttgart (thus he had a fluent knowledge of 
German), 
but he was known in the scientific community for his excellent work in 
relativity as well. The astrophysicists of the Academy planned               
further tests of Einstein's theory by astronomical means. 
Walter Grotrian, who now headed the Academy's Astrophysical Observatory at 
Potsdam, organized in collaboration with
Finley-Freundlich (at St Andrews, Scotland) and with Papapetrou's and 
even  Einstein's\footnote {An exchange of  letters between Naas and Einstein 
from November 1951 deals with the scientific chances of such a campaign  
\cite{EinsteinNaas}.} 
advice a campaign to
observe the total Solar eclipse on June 30, 1954 from the Svedish island
\"Oland. As for many earlier expeditions, cloudy sky prevented any
observation. 

Apparently, these efforts stimulated Papapetrou's research interests. 
Already his latest papers in Manchester were concerned with the derivation
of equations of motion for spinning test particles from the conservation
law ${\cal T}^{\mu\nu}_{;\nu}=0$  \cite{Papapetrou52a}, \cite{Corinaldesi}. 
Removing the restriction to test bodies leads to the question how the 
gravitational field of a {\it single} spinning mass would look. For 
the nonrotating spherical point mass Schwarzschild had given the answer,
but the spinning counterpart was still an open question, at least as far as 
an exact solution was concerned. Solving  this problem was of principle 
interest for tests of General Relativity in the Solar system 
with the rotating Sun and planets, no matter how small the effects of rotation
would turn out finally. 
       
Problems of this type had already attracted several theoreticians.  
As early as 1918 J. Lense and H. Thirring in Vienna had calculated the 
exterior gravitational field of a rotating sphere, describing the influence 
of rotation as linear perturbation to the Schwarzschild metric 
\cite{Lense}. R. Bach \cite{Bach} continued the Lense-Thirring calculation  
by adding terms which are quadratic in the rotation velocity. In 1924 K. 
Lanczos published a simple exact solution of the
matter field equations for  uniformly rotating dust 
\cite{Lanczos}. The matter density in his model has a minimum on the rotation 
axis and increases exponentially    
with the coordinate distance             
from the axis, thus compensating the 
increasing centrifugal forces by an increased gravitational attraction.
Later papers by W.R. Andress \cite{Andress} and E.S. Akeley 
\cite{Akeley1,Akeley2} 
were mainly concerned with approximation methods for axisymmetric 
stationary fields. The first exact 
solutions of the vacuum field equations within this class of fields 
were found by T. Lewis and published in an important paper in 1932 
\cite{Lewis}. A few years later  
W.J. Van Stockum \cite{Stockum} derived the gravitational field of an infinite 
rotating cylinder of dust particles and used one of the Lewis 
solutions to fit this interior field to an exterior vacuum field.
A different class of exact vacuum solution of the Lewis equations 
was given by Papapetrou soon after he arrived in 
Berlin \cite{Papapetrou53}.                              
Exact solutions including time-independent 
gravitational fields were also systematically studied by
Pascual Jordan's research group at Hamburg university,  
mainly by J. Ehlers, W. Kundt and E. Sch\"ucking 
\cite{Jordan},\cite{Witten}.  

However, no exact solution discovered so far could be considered as the 
gravitational field of a rotating nearly spherical mass. 
All of them had either singularities on the axis interpreted as violation of 
the vacuum equations and presence of a line distribution of rotating matter
or, as in Papapetrou's 1953 solutions \cite{Papapetrou53}, had angular momentum
but zero total mass, as measured by the corresponding terms in an asymptotic
expansion of the metric.

From a geometrical point of view, these vacuum 
solutions did belong to a class
of metrics, which were later shown by Papapetrou \cite{Papapetrou64} as 
invariantly characterized by the existence of two commuting Killing fields 
$\xi^\rho$ (timelike) and $\eta^\rho$ (spacelike with closed orbits), 
which admit 2-spaces orthogonal to the group orbits. Usually this class of 
(vacuum) solutions is called the Lewis-Papapetrou class - also the Kerr metric 
belongs to this class.
 
The author, 
who had a background in astrophysics from Schwarzschild's Potsdam Observatory,
entered Papapetrou's small research group in the mathematics 
institute in May 1959.
I enjoyed the stimulating atmosphere with regular guests from 
East German universities and international visitors like    
Marie-Antoinette Tonnelat from Paris and Felix Pirani from London. 
My first duty was to solve a problem in Einstein's         
field theory with the 
asymmetric metric tensor. Having stood this test \cite{Daut59}, not without
help by Papapetrou, he considered me as being able to treat 
a more complicated problem.
"Find the gravitational field of a rotating 
point mass as a suitable generalization of the Schwarzschild metric" was 
his suggestion in the late autumn of 1959.
Unfortunately for the project, Papapetrou was invited to visit the 
Institute Henri Poincar\'{e} in Paris for one year. He left Berlin early in 
1960, thus his valuable advice and encouragement was missed. After all, 
communication in that time without electronic mail 
was mainly confined to sending yellow post letters 
occasionally. 

In the next sections this old attempt to tackle the problem is described.
In the final section I return to the further history of the Kerr field.            
\section{\label{sec:level2} Lewis equations and their generalization}
The immense literature on stationary axisymmetric
gravitational fields known today \cite{ExactSol03} did not exist in 1959, 
apart from the
three basic papers by Lewis, van Stockum and Papapetrou. To these papers 
one should have added J\"urgen Ehlers' 1957 thesis
on exact solutions \cite{Ehlers57}, 
but I was only later aware  of this work.
At some time I had access to the useful Jordan report 
\cite{Jordan}, which summarized the research of the Hamburg group and included 
a chapter on stationary gravitational fields.
Apart from these papers, the whole field was unexplored territory.

The most influential paper was that of  
T. Lewis \cite{Lewis}. He wrote the line element without further explanation
essentially as (the notation is taken from Papapetrou \cite{Papapetrou53}) 
\be
ds^2 =  e^\mu(dx_1^2 + dx_2^2)  +l d\phi^2 +2 m d\phi dt -f dt^2,
\ee
where all functions $\mu,l,m$ and $f$ depend only on the two coordinates
$x_1,x_2$.
As shown by Lewis, with his metric the vacuum field equations $R_{\mu\nu}=0$ 
have a clear 
structure and admit a straightforward integration procedure: One obtains a 
set of three coupled nonlinear partial differential 
equations in two dimensions involving only the three functions $f,l,m$. 
Further equations
allow to determine the remaining function $\mu$ by simple integration, 
provided a solution $f,l,m$ is given. 

Lewis and Papapetrou had found special classes of solutions, but not yet one
which was singular only along a single worldline and had a specific 
asymptotic behaviour, tending to Minkowski spacetime at spatial infinity.

I began to treat the problem in a systematic way. As it
turned out, this was not the best method. 
The first question was: Is the form of the metric tensor {\it assumed} by
Lewis already general enough to describe that axisymmetric
vacuum field, which we wanted to find? Actually Lewis served well,
but this could not be known beforehand.  

We had always assumed the metric field to 
admit two {\it commuting} Killing vectors 
$\xi^\rho$ (timelike) and $\eta^\rho$ (spacelike, at least near the axis), thus 
$\xi^\rho_{;\sigma}\eta^\sigma- \eta^\rho_{;\sigma}\xi^\sigma =0$.
One should also have asked if this assumption of an {\emph Abelian}
isometry group $ G_2$ is perhaps a restriction for the problem. 
I do not remember our arguments for adopting commutativity (apart from Ockham's
razor). 
Ten years later Brandon Carter \cite{Carter70} proved that the 
commutativity assumption means no loss of generality:              
Axisymmetric stationary fields which become asymptotically flat 
have commuting Killing fields. More recently Alan Barnes 
\cite{Barnes} noted that the 
Abelian character of $G_2$ follows in a simple manner from the fact that the 
orbits of $\eta^\rho$ should be topologically circles. 
In any case, writing down normal 
forms for the metric in the case of a non-Abelian $G_2$ would have       
convinced us that the running coordinate along the orbits of 
$\eta^\rho$ could not be a cyclic one. 

Then, asssuming an Abelian $G_2$, it was easy to see that one can introduce 
new coordinates by 
requiring  $\xi^\rho =
 \delta^\rho_0, \eta^\rho = \delta^\rho_3$, such that the metric depends only 
on the coordinates $x^1,x^2$. This special coordinate form for the Killing 
fields is left invariant by coordinate transformations of the type 
\be
\bar{x}^1 = \bar{x}^1(x^1,x^2),~\bar{x}^2 = \bar{x}^2(x^1,x^2), \label{tr1}
\ee
\be
\bar{x}^3 = x^3 + p(x^1,x^2),~\bar{x}^0 = x^0 + q(x^1,x^2), \label{tr2}
\ee
apart from linear transformations of $x^3,x^0$;
$\bar{x}^1,\bar{x}^2$ are 
invertible and $p,q$ arbitrary functions of $x^1,x^2$.

The question was now: 
Is it possible to reduce 
the general axisymmetric stationary vacuum field 
to the Lewis form by means of these transformations? 
To my surprise, the answer turned out to be {\it no,
not in general}.
I tried several ways to simplify the metric. 
Successful was a sort of covariant reduction, where 
the field equations are written as three-dimensional and in a second step as
 two-dimensional covariant 
relations.  Methods of this type are described in the Jordan report 
\cite{Jordan}  and also in the Landau-Lifschitz volume "Field Theory" 
\cite{LL},
which I just translated at that time from Russian into German.
In the first step the metric tensor $g_{\mu\nu}$ was split into
\be
g_{00}=-V^2,~ g_{0i} = -\gamma_{i} V^2,~ 
g_{ik}= \gamma_{ik}-   \gamma_{i}\gamma_{k}V^2     
\ee
($i,k=1,2,3$). The three field equations $R^i_0=0$ then led to 
\be
(V^3\kappa^{ik})_{|k}=0, \label{r0i} 
\ee  
where $\kappa_{ik} \equiv \frac{1}{2} (\gamma_{i,k}-\gamma_{k,i})$, the 
stroke denotes the 
covariant derivative with repect to the 3-metric $\gamma_{ik}$ and indices 
are muved using  $\gamma_{ik}$. If (\ref{r0i}) holds,
the quantity $E_{ikl}\kappa^{ik}V^3$, constructed with the 
three-dimensional totally antisymmetric 
Levi-Civita tensor $E_{ikl}$, must be a gradient $\psi_{,l}$.              
Solving for $\kappa_{ik}$ resulted in                                  
\be
\kappa_{ik} = \epsilon_{ikl}\gamma^{lm}\psi_{,m}/(2V^3).
\ee
The condition that $\kappa_{ik}$ is a rotation, requires that $\psi$ satisfies
the field equation 
\be
\gamma^{kl}\psi_{,k|l} - 3V_{,k}\psi_{,l}\gamma^{kl}/V=0. \label{psieq}
\ee
The other field equations $R^0_0=0$ and $R^i_k=0$ became
\be
V^3\gamma^{kl}V_{,k|l} + 2\psi_{,k}\psi_{,l}\gamma^{kl}  =0, \label{3r00}
\ee
\be
R^{(3)i}_k -\frac{1}{V}\gamma^{il}V_{,k|l} 
+\frac{2}{V^4}\delta^i_k\psi_{,l}\psi_{,m}\gamma^{lm}  
-\frac{2}{V^4}\psi_{,k}\psi_{,l}\gamma^{il} = 0. \label{3rik}
\ee

In a second step, the 3-metric $\gamma_{ik}$ was split into two-dimensional
covariant quantities  (capital indices  always take values 1,2 in this
article):
\be
 \gamma_{33}=W^2,
~\gamma_{3A}=\epsilon_A W^2,
~\gamma_{AB}=\epsilon_{AB}+ \epsilon{_A}\epsilon{_B}W^2.  
\ee
The 3-tensor equation (\ref{3rik}) split into a scalar, vector and tensor
equation in two dimensions:
\be \fl \qquad 
\epsilon^{AB}W_{,A\parallel B}/W - W^2 k^{AB}k_{AB} 
- 2\psi_{,A}\psi_{,B}\epsilon^{AB}/V^4 +W_{,A}V_{,B}\epsilon^{AB} /(V W) =0,
\label{2r33}
\ee
\be \fl \qquad 
(W^3k^{AB})_{\parallel B} +  W^3k^{AB}V_{,B}/V =0, \label{2r3A} 
\ee
\ben \fl \qquad   
R^{(2)A}_B -\epsilon^{AC}W_{,B\parallel C}/W -2W^2 k^{AC}k_{BC}
-\epsilon^{AC}V_{,B\parallel C}/V  \nonumber
\een
\be \fl \qquad        
+2\delta^A_B \psi_{,C}\psi_{,D}\epsilon^{CD}/V^4
-2\psi_{,B}\psi{,_C}\epsilon^{AC}/V^4 =0.
    \label{2rAB}
\ee
Here 
\be
k_{AB} = \frac{1}{2}(\epsilon_{A,B}-\epsilon_{B,A})
\ee
and the double stroke denotes the covariant derivative with respect to the 
Christoffel affinity formed with the 2-metric $\epsilon_{AB}$ (indices
are moved using  $\epsilon_{AB})$.
Using the coordinate transformations (\ref{tr1}), I assumed that the 2-metric 
can be transformed into a conformally flat metric:                             
\be
\epsilon_{AB} =  \delta_{AB} e^{\mu}. \label{2m}
\ee
The equations (\ref{2r3A}) now became explicitly     
\be
k_{AB,B}+ 3 k_{AB}W_{,B}/W+k_{AB}V_{,B}/V - k_{AB}\mu_{,B} =0. 
\ee 
The integration gave             
\be
k_{12} = \frac{k e^{\mu}}{VW^3} \label{keq}
\ee
with  $k$ as integration constant. The relations (\ref{2m}) and (\ref{keq})
simplified the field equations considerably. The two equations (\ref{psieq}) 
and (\ref{3r00}) became
\be
 \fl \qquad 
\Delta\psi - [\ln{\frac{V^3}{W}},\psi] = 0, \label{psi}
\ee
\be
\fl \qquad \Delta V +  [V,W]/W  +2 [\psi,\psi]/V^3 = 0,
\label{V}
\ee
where  the differential operator $\Delta$ is the Laplacian in two dimensions, 
$\Delta = \frac{\partial^2}{\partial x_1^2}
+\frac{\partial^2}{\partial x_2^2}$,
and the Lewis bracket is defined as
\be [A,B]  \equiv  \frac{\partial A}{\partial x^1 }
\frac{\partial B}{\partial x^1 }+
  \frac{\partial A}{\partial x^2}\frac{\partial B}{\partial x^2}.  
\ee
Similarly, (\ref{2r33}) is a differential equation for $W$:  
\be
\fl \qquad \frac{\Delta W}{W}+ \frac{[V,W]}{VW} -2 \frac{[\psi,\psi]}{V^4} = 
       \frac{2k^2e^{\mu}}{V^2W^4}.  \label{W}                          
\ee
The remaining equations (\ref{2rAB}) were              
\be
\fl \qquad 
\Delta\mu +\frac{\Delta W}{W}+ \frac{\Delta V}{V}
                       -2\frac{[\psi,\psi]}{V^4}
=    -\frac{4k^2e^{\mu}}{V^2W^4},  \label{omega}
\ee
\be
\fl \qquad
  \mu_{,1}(\frac{V_{,1}}{V}+ \frac{W_{,1}}{W})
- \mu_{,2}(\frac{V_{,2}}{V}+ \frac{W_{,2}}{W})
=   2\frac{\psi_{,1}^2}{V^4}-2\frac{\psi_{,2}^2}{V^4}+\frac{W_{,11}}W{} 
-\frac{W_{,22}}W{}
 +\frac{V_{,11}}V{} -\frac{V_{,22}}{V}   , \label{om1}
\ee
\be
\fl \qquad
\mu_{,1}(\frac{V_{,2}}{V}+ \frac{W_{,2}}{W}) +   
 \mu_{,2} (\frac{V_{,1}}{V}+ \frac{W_{,1}}{W}) =   
2\frac{V_{,12}}{V} +2\frac{W_{,12}}{W}+ 4 \frac{\psi_{,1}\psi_{,2}}{V^4}. 
\label{om2}   
\ee
For the function $R = V W$ one derives the simple relation
\be
R^3 \Delta R = 2k^2V^2 e^{\mu}. \label{R}
\ee
Apparently, Lewis was not general enough.
The field equations 
(\ref{psi},\ref{V},\ref{W}-\ref{R}) 
differ from the Lewis equations through the
occurence of an integration constant $k$, complicating Lewis' integration 
scheme (note however, $\mu_{,1}$ and $\mu_{,2}$ as calculated from 
(\ref{om1},\ref{om2})
still satisfy $\mu_{,1,2}= \mu_{,2,1}$ as well as (\ref{omega}) in 
virtue of the other equations, even if $k \neq 0$).
The complication was not the only problem. I had also carried out the reduction 
process inversely, splitting the metric first with regard to $\eta^\mu$
and then to $\xi^\mu$. This  introduced a {\it different} set of equations 
with a {\it new} constant $ \bar{k}$. But neither set could represent the 
{\it full} 
system of vacuum equations in the Abelian $G_2$ case, since {\it both} 
constants $k,\bar{k}$ are expected to occur. The reason why the derivation 
given above missed $\bar{k}$ is an implicit assumption made in the 
calculation of (\ref{psieq})-(\ref{3rik}), that the potential $\psi$ like all 
other functions does
not depend on $x^3$, i.e. ${\cal L}_\eta \psi =0$. Yet the condition  
${\cal L}_\eta \psi = -2\bar{k} \neq 0$ 
is compatible with the Killing symmetries for the metric and leads to 
the complete system of vacuum field equations in the presence of two commuting 
Killing fields.

Today one recognizes that the constants 
$k$ and $\bar{k}$ are essentially the twist scalars associated with the 
two Killing vectors $\eta^\mu, \xi^\mu$:
\be
2k = E_{\mu\nu\rho\sigma}\xi^\mu \eta^\nu \eta^{\rho ; \sigma},~
2\bar{k} = E_{\mu\nu\rho\sigma}\eta^\mu \xi^\nu \xi^{\rho ; \sigma}. 
\label{kk}   
\ee
We know that $k$ and $\bar{k}$ are constants if the 
vacuum equations hold, more generally, they are constants if and 
only if the conditions
\be
E^{\mu\nu\rho\sigma}\xi^\lambda  R_{\lambda\mu}\xi_\nu \eta_\rho = 
E^{\mu\nu\rho\sigma}\eta^\lambda  R_{\lambda\mu}\eta_\nu \xi_\rho =0
               \label{circ}     
\ee   
are satisfied \cite{ExactSol03},\cite{Wald}.

Little is known about the existence of solutions of the generalized Lewis 
equations.  Apparently even today no vacuum solution is known 
(cf. \cite{ExactSol03}, see however \cite{Gaffet}). R. Geroch has shown 
that for $k,\bar{k} \neq 0$ 
his method of generating new vacuum solutions from given ones 
breaks down in the sense that the presence of Killing fields is not 
preserved \cite{Geroch72}, \cite{Geroch71}.         

I believe, Papapetrou was not happy with the extension of the Lewis 
equations.                                                                  
Indeed, his intuition turned out to be correct. He showed in a remarkable paper
six years later \cite{Papapetrou66}, that the {\it twist scalars must 
vanish, restoring the Lewis equations for our problem}.
The geometrical background of this result became clear in papers by
W. Kundt and M. Tr\"umper \cite{Kundt66}                                    
and by B. Carter \cite{Carter69}:
While the orbits of the two Killing vectors $\xi^\mu, \eta^\mu$ are always
two-surface forming, the two-surface elements orthogonal to the group
orbits do not fit to finite surfaces for non-vanishing twist scalars. 
The Lewis block diagonal form
of the metric is  just equivalent to "orthogonal transitivity", to
the existence of two-surfaces orthogonal to the group surfaces.

Papapetrou's 1966 result could have been        
found already in 1960, had the boundary 
conditions on the symmetry axis been analyzed:  We had fairly precise 
ideas for 
the behaviour of the metric at spatial infinity, but did not consider 
the axis, since here unknown  singularities were expected.
However, for a rotating localized mass the metric on 
 part of the axis outside the body must be regular. The existence of a
(at least partly) regular axis means that the cyclic Killing vector 
$\eta^\mu$ {\it vanishes there}       
(for a recent careful discussion of the axis conditions in axisymmetric 
spacetimes see, e.g., \cite{Rinne}). 
Since our coordinates were restricted such that $\eta^\mu = \delta^\mu_0$ 
everywhere, they must be singular on the axis. A look at 
(\ref{kk}) with the rhs now written in {\it regular} coordinates shows 
immediately 
that on the axis (and, since $k,\bar{k}$ are constants, everywhere) 
$k=\bar{k} =0$. If vacuum solutions with $k,\bar{k} \neq 0$ exist, they 
would carry singularities on the whole symmetry axis and could not represent
the exterior gravitational field of a compact body. But this was not 
recognized in 1960.
    
\section{Lewis-Papapetrou class of vacuum fields}
After wasting some time with a fruitless study of the 
extended equations, I returned to 
Lewis.  Letting $k=0$ in (\ref{R}) gives $\Delta R=0$, thus $R$ is a harmonic 
function. Provided $R$ is not a constant,
this allowed to introduce canonical coordinates $\rho=R$ and
eliminated $W=\rho/V$ in all equations.
The basic system consisted now of only two coupled nonlinear partial 
differential equations                    
for the two potentials $\psi(\rho,z)$ and  $V(\rho,z)$, depending on the
two cylindrical coordinates $x^1=\rho, x^2=z$: 

\ba  \Delta\psi + \frac{1}{\rho}\frac{\partial\psi}{\partial\rho}
 & = & \frac{4}{V}[V,\psi], \label{e1}  \\
     \Delta V + \frac{1}{\rho}\frac{\partial V}{\partial\rho}
 & = & -\frac{2}{V^3}[\psi,\psi] +\frac{1}{V}[V,V]. \label{e2}
\ea

One was faced with the problem to find exact solutions of this system with a 
prescribed behaviour at spatial infinity.                   
Lewis and Papapetrou had derived equivalent systems of 
equations  for a different set of field quantities. 
No systematic integration theory was known for either system. The interesting
mathematical properties of (\ref{e1},\ref{e2}) as a completely
integrable system were unknown at that time. 
To find solutions at all, Lewis and 
Papapetrou had to make special ad-hoc assumptions for their potentials.

Could similar assumptions be tried for the system (\ref{e1},\ref{e2})?         
Some suggestions came from an article by B. Kent Harrison, just published  
in the December 1959 issue of Physical Review \cite{Harrison}.   
He presented many exact solutions of the vacuum field equations, 
obtained with heuristic methods such as separation of variables.  
The hope was that some of these techniques, perhaps in combination, would 
work also here.         
To have greater flexibility, I first transformed the system 
(\ref{e1}),(\ref{e2}) into a more general form,  by substituting 
\be
\psi= \psi(X^1,X^2),~V= V(X^1,X^2), \label{trans}
\ee
assuming the $\psi,V$ are at least twice differentiable functions of 
$X^A$, with nonvanishing functional determinant
\be
  D \equiv \left | \begin{array} {cc} \frac{\partial\psi}{\partial X^1} &
 \frac{\partial\psi}{\partial X^2}\\ 
 \frac{\partial V}{\partial X^1} &
 \frac{\partial V}{\partial X^2} \end{array} \right | \neq 0.
\ee
Then the inverse functions $X^A = X^A(\psi,V)$ exist. 
Introducing (\ref{trans}) into (\ref{e1}),(\ref{e2}), one obtains 
equations of the type (summation convention for repeated indices)
\be
\Delta X^A + \frac{1}{\rho}\frac{\partial X^A}{\partial \rho}
+ \lambda^A_{BC}[X^B,X^C]=0, \label{e3} \label{general}
\ee
where the six quantities $\lambda^A_{BC}= \lambda^A_{CB}$ are functions
of the new independent field quantities $X^A$.  
 I observed that equation (\ref{e3}) is invariant with respect to 
substitutions
\be
   \bar{X}^A = \bar{X}^A(X^B), \label{change}
\ee
provided  the $\lambda^A_{BC}$ transform as an affine connection
(considering the $X^A$ as independent variables):
\be
\bar{\lambda}^A_{BC}   =  \frac{\p \bar{X}^A}{\p X^I}
(\frac{\p^2X^I}{\p \bar{X}^B\p \bar{X}^C} + 
\lambda^I_{DE}\frac{\p X^D}{\p \bar{X}^B}\frac{\p X^E}{\p \bar{X}^C}).   
\label{lambdatrans}     
\ee

Was there any hope to reduce                 
(\ref{e1},\ref{e2}) to {\it linear}
equations using a suitable nonlinear transformation (\ref{change})? A necessary
condition is the vanishing of the Ricci tensor                     
${\cal R}_{AB} = \lambda^C_{AB,C}-\lambda^C_{CA,B}+\lambda^C_{CD}\lambda^D_{AB}
-\lambda^C_{DA}\lambda^D_{CB}$,
formed with the $\lambda$-connection in the two-dimensional space of  
potentials $(X^1,X^2)$.
A short calculation had shown that already ${\cal R}_{11}   =-3/V^2 $ 
is nonzero, the nonlinearity could not be removed. 

The simple potential space formalism allowed to anwer also other questions.
The integration idea was to try heuristic methods                         
for {\it other} potentials, if the {\it original} 
set ($V, \psi $) failed. At least for Papapetrou's basic assumption in 
 \cite{Papapetrou53}  this hope had to be given up.              
In my notation his condition is 
\be
 V_{,1}\psi_{,2} -  V_{,2}\psi_{,1}= 0. \label{curl}
\ee        
This relation is invariant with respect to arbitrary coordinate 
transformations $x'^A  = x'^A(x^B)$, but it is also invariant with 
respect to arbitrary substitutions (\ref{change}) of the potentials.         
Thus choosing other potentials does not increase the chance to find
solutions beyond the special Papapetrou class.
Another heuristic assumption, introduced by Lewis for a pair of his 
variables, was $X^1= p(X^2)$, a functional relationship between two
potentials.
It is seen that  (\ref{curl}) is satisfied in this case,               
hence also this restriction leads only to solutions within the 
special Papapetrou class.                     

An important restriction for the solutions is the proper behaviour at 
spatial infinity, assumed as ($r=\sqrt{\rho^2+z^2}$)
\be
f \rightarrow 1,~l\rightarrow \rho^2, ~m\rightarrow 
\lambda\frac{\rho^2}{r^3}
\ee
for the Papapetrou functions $f,l,m$, where $\lambda$ is proportional to 
the angular momentum ("strong boundary condition"). For the potential $\psi$
this transforms to $\psi \rightarrow \lambda z/(2r^3)$. The strong boundary 
condition ensures that the metric tends to the Minkowski spacetime at
spatial infinity $r\rightarrow \infty$, it also provides finite values for 
the total angular momentum 
\cite{Daut61}. Unfortunately, metrics of the special Papapetrou class
which satisfy the strong boundary condition have zero total mass or energy, 
this follows immediately from an $1/r$ expansion of (\ref{e1},\ref{e2})  
and  (\ref{curl}). 

\section{Sample solutions}
Vacuum metrics with $VW=const$               
did not allow canonical coordinates, but satisfy simple equations.
One  obtains from (\ref{psi},\ref{V},\ref{W}-\ref{R}) with $k=0$ 
the compatible set
\be
\Delta\psi =0, ~~ \Delta \mu =0,~~ (V^2)_{,1}= 2\epsilon \psi_{,2},~
 (V^2)_{,2}= -2\epsilon \psi_{,1}  \label{hoff} 
\ee
($\epsilon^2=1$).
$\psi$ and $\mu$ are harmonic functions, their singularities had to be 
considered as resulting from a singular matter distribution.
 
It seems obvious here to interpret $x^1,x^2$ as 
quasi-Cartesian coordinates in a plane orthogonal to the $x^3$-axis. 
The Killing 
vector $\eta^\mu$ then represents a translational symmetry along the 
$x^3$-axis. 
The simplest solutions have a singularity at the origin of $x^1,x^2$
and were 
believed to describe the exterior gravitational field  of an 
infinite rotating cylinder along the $x^3$-axis.  
But such fields had nothing to do with              
the gravitational field of a rotating point mass. 
Later R.B. Hoffman discussed this class of stationary fields \cite{Hoffman}. 

Harrison's separation technique as applied to (\ref{general})
was my main working tool.                             
Similar methods are still used today \cite{Gariel}.
In principle, this technique can be applied not only to transformed 
potentials but also in the case of transformed coordinates.
Thus I filled many sheets of        
paper with formulae of that type, too often stopping the calculation 
once it became clear that the required strong boundary conditions could not be 
satisfied. 
 
For example, in the case of quasi-spherical coordinates 
$\rho= r \sin{\theta},~ z= r\cos{\theta}$ one has with $f=V^2$                  
\be
\psi_{,rr} +\frac{2}{r}\psi_{,r} + \frac{\psi_{,\theta\theta}}{r^2} + 
\cot{\theta}\frac{\psi_{,\theta}}{r^2} = 2\frac{f_{,r}}{f}\psi_{,r} +
\frac{2}{r^2}\frac{f_{,\theta}}{f}  \psi_{,\theta}, \label{ps}
\ee  
\be
f_{,rr} +\frac{2}{r}f_{,r} -\frac{1}{f}f_{,r}^2  
+\frac{1}{r^2}f_{,\theta\theta} 
+       \cot{\theta}\frac{f_{,\theta}}{r^2} 
-\frac{1}{r^2}\frac{f_{,\theta}^2}{f}= 
-\frac{4}{f} \psi_{,r}^2 -\frac{4}{r^2f}\psi_{,\theta}^2. \label{ff}
\ee
Separation assumptions led to a number of subcases. 
In the case where both            
$\psi$ and $f$ depend only on the radial coordinate $r$, (\ref{ps})
gives $\psi_{,r} = c f^2/r^2$, $c=const$. Introducing this into (\ref{ff})        
leads to an equation for $f$ alone:
\be
f_{,rr} + \frac{2}{r}f_{,r} - \frac{1}{f}f_{,r}^2 + 4 \frac{c^2}{r^4}f^3= 0.
\ee
The solutions are        
\be
-g_{00}= V^2 = \frac{\alpha}{\frak{Cos}(\gamma+\beta/r)},~~ 
\psi = -\frac{\alpha}{2} {\frak{Tan}}(\gamma+\beta/r).
\ee
The three integration constants $\alpha,\beta,\gamma$ ($c=\beta/(2\alpha)$) 
are not independent, requiring $lim~g_{00} \rightarrow -1 for~r\rightarrow 
\infty$ gives $\alpha= {\frak{Cos}}{\gamma}$.

Also the equations for $\mu$ can be integrated,                      
the further non-vanishing components of the metric are    
\ba
g_{rr} &=& \frac{1}{\alpha}\frak{Cos}(\gamma +\beta/r)
e^{ -\beta^2 \sin^2\theta/(4r^2)}, \\                     
g_{\theta\theta} &=& r^2g_{rr}, \\
g_{\phi\phi} &=& 
({\mathfrak{Cos}}(\gamma+\beta/r)/\alpha 
 - \frac{\beta^2 \cot^2\theta}{\alpha r^2\mathfrak{Cos}(\beta/r+\gamma)})
 r^2\sin^2{\theta},      \\
g_{\phi t} &=&   \frac{\beta \cos\theta} {2 \mathfrak{Cos}(\gamma +\beta/r)}. 
         \label{sol1}
\ea

Expansion of $g_{00}$ in reciprocal powers of $r$ shows that 
$2M= \beta~\frak{Tan}{\gamma}$ is the coefficient of the $1/r$-term. 
Several arguments suggest that $M$ is the total 
energy of the field. In 1960 I used pseudotensors derived from the Lagrange 
density of the gravitational field according to Noether's procedure 
\cite{Daut61}, but  
one obtains the same result with the ADM mass formula \cite{ADM}.
Again however,
this was clearly  not the solution we had looked for, since the boundary 
conditions at spatial infinity are not satisfied for the nondiagonal term, 
$\psi$ does not vanish but tends to the constant $-\alpha 
\frak{Tan}{\gamma}/2$ at spatial infinity.
Thus the total angular momentum \cite{Daut61} diverges.
Since $\psi,V$ depend on $r$ only,
a short look shows that (\ref{curl}) is satisfied, 
thus the solution belongs to the special Papapetrou class. The metric
has nonzero total energy only because the strong boundary conditions 
are violated.

It was time consuming and unsatisfactory to search for solutions  
with the rather simple trial-and-error methods at hand.
According to my 1960 notebook, I was impressed by Buchdahl's procedure 
to obtain new 
stationary solutions from a given static or stationary vacuum field.
Later many successful recipes were developed to realize this idea
of solution generation \cite{Kordas},\cite{ExactSol03},  
starting from the pioneering papers by Buchdahl 
 \cite{Buchdahl1}, \cite{Buchdahl2} and by Ehlers \cite{Ehlers59}. 
This has finally opened the door to the solution space.                                   
\section{Any chance in 1960? }

But still I did not give up. 
On March 6, 1960, I sent Papapetrou a letter with a short summary of       
results obtained so far (translated from German, signature of metric changed):

\baselineskip=12pt
\begin{quote}
{\it   
``[...] After two weeks of skiing with best snow conditions I'm back to 
Berlin.  The state of my work is roughly as follows.
\smallskip

1) It was guessed that forming a normal form for $g_{\mu\nu}$ already before
the field equations are reduced is preferable, since more transformation 
freedom is available. However, Petrov's choice ($g_{11}= 1,g_{12}=g_{13}=
g_{10}=0$) offers no advantage to my previous approach for a concrete solution 
of the field equations. It stands for a certain choice of coordinates in the
$x_1$-$x_2$-space, but the field equations which must be solved next are not 
simplified.
\smallskip

2) The reverse reduction method (first $x_3$ then $x_0$) gives very 
complicated equations as in the previous case, if $\bar{k}\neq 0$. ($\bar{k}=0$
now means $g_{0I}=g_{I3}g_{03}/g_{33}$; for the original reduction sequence
the analogous condition $k=0$ was $g_{3I}=g_{0I}g_{03}/g_{00}$). But if
we assume $\bar{k}=0$ and require Minkowskian boundary conditions, the 
equations reduce to those of Lewis. 
\smallskip

3) A class of solutions ($k=0$), which are presumably uninteresting physically,
obviously describe the gravitational field of a rotating cylinder with 
a multipole matter source. These fields are independent of $z$, but they 
lost rotational symmetry, depending on the polar coordinates in a plane
orthogonal to the cylinder axis. A closer inspection seems not to be 
worthwhile. 
\smallskip

4) My earlier solution}   \end{quote}
\vspace*{.1cm}
\begin{eqnarray*} 
ds^2= -\frac{\alpha}{\frak{Cos}(\beta/r+\gamma)}dt^2 +
\frac{\frak{Cos}(\beta/r+\gamma)}{\alpha}
 e^{ -\beta^2 \sin^2\theta/(4r^2)}(dr^2+r^2d\theta^2) \\ 
+(\frac{\frak{Cos}(\beta/r+\gamma)}{\alpha} 
- \frac{\beta^2 \cot^2\theta}{\alpha r^2\frak{Cos}(\beta/r+\gamma)})
r^2\sin^2{\theta}d\phi^2 
+\frac{\beta \cos\theta} {\frak{Cos}(\beta/r+\gamma)}d\phi dt
\end{eqnarray*}
\vspace*{.1cm}
\begin{quote}  
{\it has some interesting properties which might render it acceptable 
in spite of the missing boundary condition for $g_{03}$:

{\parindent0.5cm

(a) The total energy (according to M\o ller) is finite and
    depends as in the Schwarzschild case on the $1/r$ term in the expansion of 
    $g_{00}$.

(b) The total momentum vanishes.

(c) For $\alpha \rightarrow \infty$ and neglecting the terms with $1/r^2$ 
    the solution tends asymptotically to the Schwarzschild solution.

(d) For $r \rightarrow \infty$ the space becomes homogeneous, but is no longer 
    isotropic. Since there exists a distinguished direction, this appears to
    be reasonable.}

Mr. Treder and I agree that this type of solution (the given one 
is only the simplest) should be considered as physically reasonable. 
I would like to ask you for your opinion.
\smallskip

5) Presently I am trying to transform the field equations with $k=0$ using 
suitable coordinate conditions in the $x_1$-$x_2$-space to give them a 
convenient structure. I hope that within one of these coordinate systems the 
solution 
which we are looking for takes a fairly simple form, and can therefore be 
found relatively easily. But one must be lucky! 

In spite of the small success so far I still believe that one can find a 
stationary solution which satisfies all requirements.} 
  '' \end{quote}

\baselineskip=14pt

It would have been interesting to know Papapetrou's 
reaction, but I do not remember having obtained a response.                   

Needless to say, I had no luck with the recipe proposed under item 5).
But the  recipe itself  - looking for suitable coordinates in the  
$x^1$-$x^2$-space - 
was indeed a route to the Holy Grail, the rotating Schwarzschild field: 
 Transforming the cylindrical coordinates 
$\rho,z$ into some kind of radial and angular coordinates $r,\theta$ 
by means of 

\be
\rho= \sqrt{r^2+a^2-2rM\sin{\theta}},~ z= (r-M)\cos{\theta},  \label{blt}
\ee
as done by F.J. Ernst \cite{Ernst68}  after Kerr's discovery,
leads to the Kerr solution with the potentials 
\be
V^2 = 1-\frac{2rM}{r^2+a^2\cos^2{\theta}},~~  
\psi = -\frac{aM\cos{\theta}}{r^2+a^2\cos^2{\theta}}. \label{k1}
\ee
This looks simple indeed. But how could one have
figured out the  coordinate transformation (\ref{blt}) in 1960? 
There exist other coordinates in which the Kerr functions appear fairly simple,
e.g. spheroidal prolate coordinates \cite{Bergamini}.
To find them by trial-and-error would not have been easy, but it   
was not impossible, given sufficient diligence and persistence.

In 1968 F.J. Ernst also found that the complex combination 
 ${\cal{E}} = V^2+ 2i \psi$ satisfies an elegant differential equation
 \cite{Ernst68}, 
 easily derivable from (\ref{e1},\ref{e2}), 
which has dominated the research on stationary axisymmetric 
gravitational fields since that time. 
The Kerr solution of the complex Ernst equation has 
the simple form $ {\cal{E}} = 1- 2M/(r-ia\cos{\theta})$ 
in the coordinates $r,\theta$  given by (\ref{blt}).
The related function $\xi = (1+ {\cal{E}})/(1-\cal{E})$ 
(essentially a potential transformation as discussed above)
satisfies a similar differential equation.                       
Kerr is then represented by $\xi = r/M  -ia\cos{\theta}/1 -1$. Evidently,
this is now even an almost {\it trivial} solution of the differential equation 
for $\xi$. A solution could hardly be simpler. 

It is satisfying that both concepts, coordinate transformation in  
$x^1$-$x^2$-space and potential transformation, 
were finally so successful, thanks to the efforts by F.J. Ernst.

\section{The winner is Kerr }
  The slow progress as well as Papapetrou's absence from Berlin 
rapidly diminished the amount of time I 
spent in 1960 for the rotating Schwarzschild problem. 
During the year I became interested in many other questions of 
gravitation. Not being guided by my boss, I worked on boundary conditions, 
surface layers, shock waves and other problems of 
gravitational radiation. In particular, the characteristic 
initial value problem for the Einstein field equations was a very 
interesting topic,
since here one could handle the two intrinsic degrees of freedom 
of the gravitational field rather directly.  
When Papapetrou returned from
Paris in the beginning of 1961, I confronted him with new ideas
about these perhaps more actual problems.                       
There was a rumour that             
Bondi and his group in London were working on similar questions.
The problem of stationary fields was forgotten for the present,
at least for me.
 
In the meantime the political situation had changed for the worse.
There was the danger that Papapetrou's position as prominent scientist in a 
communist country could lead to problems for                   
his relatives in Greece. The Cold War was present everywhere, particularly 
in the divided city of Berlin, whose borders were still open. A steadily 
increasing number of East Germans escaped to the West. 
Suddenly, on August 13, 1961, the borders were closed by the communist 
authorities. Quite unexpectedly, the Berlin Wall was built, cutting off the 
free Western part of Berlin from the surrounding East German territory.
For Papapetrou and his wife the year in Paris was so pleasing compared 
with the difficult situation in East Berlin, that they decided
 - weeks before the borders were closed - 
 to stay permanently in Paris. Not being citizens of the 
East German state, they were allowed to go.   

This was unpleasent news.             
I had to  finish my PhD thesis on 
wave solutions and the characteristic initial value problem 
sufficiently early in 1961 that Papapetrou could act as adviser,   
before he finally left East Berlin at the end of the year.         
Apparently, no time was left to discuss rotating metrics in depth. 

While I did not seriously return to the spinning body, Papapetrou never 
forgot the problem.
In at least two papers \cite{Papapetrou63}, \cite{Papapetrou64} he 
further explored the properties of axisymmetric metrics 
and found a new subclass of solutions.  
However, also these metrics did not satisfy the required boundary
conditions at infinity.

Also other groups had no luck.
In their well-known survey on exact solutions published in the legendary 
Witten volume in 1962, J. Ehlers and W. Kundt had to admit that 
{\it  ``the old problem
of constructing rigorously the field of a finite rotating body is as yet 
unsolved, even as to its exterior part''} \cite{Witten}.
 
The solution of the long-standing problem                     
came 1963 from the New Zealander Roy P. Kerr in a different way. 
Kerr received his PhD 1959 from 
Cambridge University (MA),     
worked later at Syracuse University and with the US Air Force relativity 
group under Joshua Goldberg at Wright-Patterson Field in Ohio,   
before he came 
to the University of Texas at Austin in the academic year 1962/1963. Here, 
in a newly founded Center of Relativity, organized by Alfred Schild, a 
circle of relativists had gathered, including  
besides Schild and Kerr temporarily also Roger Penrose, Ray Sachs, 
Engelbert Sch\"ucking 
and other excellent scientists.
In a recent article Kerr gave a detailed description of his 
discovery in this stimulating environment  \cite{Kerr07}. 
He used a kind of null tetrad formalism, assuming from the beginning an 
algebraically
special spacetime. Both the Schwarzschild metric as well as the Kerr
metric are of Petrov type D, thus this restriction was crucial. In rather 
complicated calculations (more complicated than I ever tried) he further
restricted the fields to satisfy stationary and then axisymmetric symmetries,
so he finally found the famous solution bearing his name. His 
two-page paper "Gravitational Field of a Spinning Mass as an Example of 
Algebraically Special Metrics"                   
was published in the September 1, 1963 issue of Physical Review Letters.

Kerr presented his solution at the First Texas symposium on Relativistic 
Astrophysics held in Dallas in December 1963 \cite{Kerr64}.
Papapetrou was possibly not aware of Kerr's article when he
came to the Texas symposium, since the note \cite{Papapetrou64} 
was presented by Louis de Broglie on December 4 (S\'{e}ance du 4 d\'{e}cembre),
apparently before his departure to the States: therein is no reference to 
Kerr's paper. Kip Thorne has given a
vivid description of the Texas meeting in his book \cite{Thorne94}:

\baselineskip=12pt
\begin{quote}
{\it
``To foster dialogue between the relativists and the astronomers and 
astrophysicists, and to catalyze progress in the study of quasars, a 
conference of three hundred scientists was held on 16-18 December 1963, 
in Dallas, Texas [...] Lectures went on almost continuously from 8:30 in the 
morning until 6 in the evening with an hour out for lunch, plus 6 P.M. until 
typically 2 A.M. for informal discussions and arguments. Slipped in among the 
lectures was a short, ten-minute presentation by a young New Zealander 
mathematician, Roy Kerr, who was unknown to the other participiants. Kerr had 
just discovered his solution of the Einstein field equation - the solution 
which, one decade later, would turn out to describe all properties of spinning 
black holes, including their storage and release of rotational energy [...] ; 
the solution which  [...] would ultimately become a foundation for explaining 
the quasars' energy. However, in 1963 Kerr's solution seemed to most 
scientists only a mathematical curiosity; nobody even knew it described a 
black hole -- though Kerr speculated it might somehow give insight into the 
implosion of rotating stars.

The astronomers and astrophysicists had come to Dallas to discuss quasars;
they were not at all interested in Kerr's esoteric mathematical topic. So, 
as Kerr got up to speak, many slipped out of the lecture hall and into the
foyer to argue with each other about their favorite theories of quasars.
Others, less polite, remained seated in the hall and argued in whispers.
Many of the rest catnapped in a fruitless effort to remedy their sleep 
deficits from the late-night science. Only a handful of relativists listened,
with rapt attention. 

This was more than Achilles Papapetrou, one of the world's leading relativists,
could stand. As Kerr finished, Papapetrou demanded the floor, stood up, and 
with deep feeling explained the importance of Kerr's feat. He, Papapetrou, had
been trying for thirty years to find such a solution of Einstein's equation,
and had failed, as had many other relativists. The astronomers and 
astrophysicists nodded politely, and then, as the next speaker began to hold 
forth on a theory of quasars, they refocused their attention, and the meeting 
picked up pace. ''  }
\end{quote}  \baselineskip=14pt
Kerr's paper of 1963 is a masterpiece of clarity and conciseness and, as     
Chandrasekhar noted, is "{\it surprisingly complete in enumerating the 
essential 
features of the solution}" \cite{Chandra}. It has only one deficiency,    
it gives few hints how the solution was derived. 
When Wolfgang Kundt came to East Berlin in the spring of 1964 for a visit, 
we discussed Kerr's metric, but found it hard to verify that 
it is indeed a solution of the vacuum equations. 
Many of our colleagues had the same problem. 

Summarizing, one can shortly answer why the pre-Kerr approaches failed:
We had the adequate differential equations,
essentially identical to the Ernst equation  
split into real and imaginary parts. 
Yet the group-theoretical properties of the solutions were not recognized,
and thus no proper key to the unexpectedly large solution space was found.
 
Kerr's new way circumvented the problem                                    
by a  restriction to algebraically special metrics                   
from the beginning, a condition which 
could not easily be expressed in the formalism we had used.
In spite of 
more complicated equations he finally
brilliantly succeeded, not least because of his persistence.

\section*{Acknowledgements}
The author is grateful to R. Kerr, W. Kundt, A. Rendall and 
G. Wallis for reading the manuscript and for comments and suggestions.

\end{document}